\documentclass[
prl%
 ,twocolumn%
 ,secnumarabic%
,amssymb, amsmath,nobibnotes, showpacs,aps, prl]{revtex4}
\usepackage{bm}%
\expandafter\ifx\csname package@font\endcsname\relax\else
 \expandafter\expandafter
 \expandafter\usepackage
 \expandafter\expandafter
 \expandafter{\csname package@font\endcsname}%
\fi
\begin{document}

\title{Standard Model Extension with Gravity and Gravitational Baryogenesis}
\author{G. Lambiase$^{b,c}$}
\affiliation{$^b$Dipartimento di Fisica "E.R. Caianiello"
 Universit\'a di Salerno, 84081 Baronissi (Sa), Italy.}
 \affiliation{$^c$INFN - Gruppo Collegato di Salerno, Italy.}

\begin{abstract}
The Standard Model Extension with the inclusion of gravity is
studied in the framework of the gravitational baryogenesis, a
mechanism to generate the baryon asymmetry based on the coupling
between the Ricci scalar curvature and the baryon current
$(\partial_\mu R)J^\mu$. We show that, during the radiation era of
the expanding Universe, a non vanishing time derivative of the
Ricci curvature arises as a consequence of the coupling between
the coefficients for the Lorentz and CPT violation and Ricci's
tensor. The order of magnitude for these coefficients are derived
from current bounds on baryon asymmetry.
\end{abstract}

 \maketitle

Studies of a possible breakdown of the fundamental symmetries in
physics have received in the last years a more and more growing
interest and have been carried out in different areas (see for
example \cite{kostelecky,ellis,loop,NCG,bertolami,mattingly}).
Referring to Lorentz's and CPT symmetries, the more general
setting in which they have been studied is the Standard Model
Extension (SME) \cite{kostelecky}. According to it, the violation
of such fundamental symmetries follows from the observation that
the vacuum solution of the theory could spontaneously violate the
Lorentz and CPT invariance, even though them are preserved by the
underlying theory. Modern tests for Lorentz and CPT invariance
breakdown have been discussed in \cite{mattinglytests}. Recently,
the SME has been extended to incorporate the gravitational
interaction. Such studies have been formalized by Kosteleck\'y and
Bluhm in the papers \cite{kostelecky,bluhm}.

This paper concerns a cosmological aspect of the SME with gravity
related to the so called {\it gravitational} baryogenesis. The
origin of the the baryon number asymmetry is an open issue of the
modern Cosmology and particle physics. Measurements of CMB
combined with the large structure of the Universe \cite{bennet},
as well as predictions of Big-Bang Nucleosynthesis \cite{burles}
give indications that matter in the Universe is dominant over
antimatter. The order of magnitude of such a asymmetry is
$\eta=\frac{n_B-n_{\bar B}}{s}\lesssim 9\,\, 10^{-11}$, where
$n_B$ ($n_{\bar B}$) is the baryon (antibaryon) number density,
and $s$ the entropy of the Universe. Conventionally, it is argued
that to generate (dynamically) the baryon asymmetry from an
initial symmetric phase the following requirements  are necessary
\cite{sakharov}: 1) baryon number processes violating in particle
interactions; 2) $C$ and $CP$ violation in order that processes
generating $B$ are more rapid with respect to $\bar{B}$; 3) out of
the equilibrium: since $m_B=m_{\bar{B}}$, as follows from $CPT$
symmetry, the equilibrium space phase density of particles and
antiparticles are the same. To maintain the number of baryon and
anti-baryon different, i.e. $n_B\neq n_{\bar{B}}$, the reaction
should freeze out before particles and antiparticles achieve the
thermodynamical equilibrium. However, a dynamically violation of
CPT allows to generate the baryon number asymmetry also in regime
of thermal equilibrium \cite{cohen}. A new mechanism to generate
the baryon number asymmetry during the expansion of the Universe
has been proposed by Davoudiasl et al. \cite{steinhardt}. In this
mechanism the thermal equilibrium is maintained and CPT (and CP)
symmetry is dynamically broken. The interaction is described by a
coupling between the derivative of the Ricci scalar curvature $R$
and the baryon current $J^\mu$ \cite{note}
\begin{equation}\label{riccicoupling}
  \frac{1}{M_*^2}\int d^4x \sqrt{-g}J^\mu\partial_\mu R\,,
\end{equation}
where $M_*$ is the cutoff scale characterizing the effective
theory. The operator (\ref{riccicoupling}) may arise from
Supergravity theories from a higher dimensional operator
\cite{kugo}.

A net baryon asymmetry can be generated in thermal equilibrium
provided that there exist interactions violating the baryon number
$B$. Such an asymmetry gets frozen-in after the decoupling
temperature $T_D$ \cite{note1}.
Since the scalar curvature only depends on cosmic time, the
effective chemical potential for baryons following from Eq.
(\ref{riccicoupling}) is $\mu_B={\dot R}/M_*^2$ (for antibaryon
one has $\mu_{\bar B}=-\mu_B$). In the regime $T\gg m_B$, the net
baryon number density at the equilibrium is given by $n_B=g_b\mu_B
T^2/6$. Here $g_b$ represents the number of intrinsic degrees of
freedom of baryons. The baryon number to entropy ratio evaluated
at the decoupling, leads to \cite{steinhardt}
\begin{equation}\label{nB/s}
  \frac{n_B}{s}\simeq -\frac{15g_g}{4\pi^2g_*}\frac{{\dot R}(t_D)}{M_*^2
  T_D}\,,
\end{equation}
where $t_D$ is the decoupling time, $s=2\pi^2g_{*s}T^3/45$, the
dot stands for the derivative with respect to the cosmic time, and
$g_{*s}$ counts the total degrees of freedom for particles that
contribute to the entropy of the Universe. The latter assumes
values close to the total degrees of freedom of effective massless
particles $g_*$ \cite{kolb} ($g_{*s}\simeq g_*\sim 106$). In order
that $n_B/s\neq 0$, a nonvanishing time derivative of the Ricci
scalar is required. Einstein's field equations imply $R=-8\pi GT_g
=-8\pi G(1-3w)\rho$, where $T_g$ is trace of the energy-momentum
tensor of matter $T_g^{\mu\nu}$, $\rho$ is the matter density,
$w=p/\rho$ with $p$ the pressure. In the radiation dominated epoch
of the standard Friedman-Robertson-Walker (FRW) cosmology, the
constant $w$ assumes the value (in the limit of exact conformal
symmetry) $w=1/3$, so that the time derivative of the Ricci scalar
is zero, as well as $n_B/s$. Differently to Einstein's theory of
gravity, a net baryon asymmetry may be generated during the
radiation era by Lorentz violating terms which couple to Ricci's
and Riemann's tensors (for other scenarios see Refs.
\cite{steinhardt,li}, and Refs. \cite{scalar} for the case in
which baryon current couples to scalar fields). Corrections
induced by Lorentz and CPT violation also affects the baryon
current $J^\mu$ in (\ref{riccicoupling}). Nevertheless the latter
can be neglected since they give rise to corrections of the second
order. We note that the effects on baryogenesis of spontaneous CPT
violation in a string inspired scenario has been studied by
Bertolami et al. \cite{bertolamiBA} (see also \cite{carroll}).

The SME with the inclusion of gravitational interactions
\cite{kosteleckyG} foresees that the effective action is
$S=S_{HE}+S_m+S_{LV}$. $S_{HE}=(16\pi G)^{-1}\int d^4x
\sqrt{-g}(R-2\Lambda)$ is the Hilbert-Einstein action of General
Relativity ($\Lambda$ is the cosmological constant), $S_m$ the
general matter action (which also includes Lorentz violating
matter gravity coupling), and finally $S_{LV}$ contains the
leading Lorentz violating gravitational couplings
\begin{equation}\label{LSaction}
S_{LV}=\frac{1}{16\pi G}\int
d^4e\left(-uR+s^{\mu\nu}R_{\mu\nu}+t^{\kappa\lambda\mu\nu}
R_{\kappa\lambda\mu\nu}\right)\,.
\end{equation}
The coefficients $u$, $s^{\mu\nu}$ and $t^{\kappa\lambda\mu\nu}$
are real and dimensionless. Moreover $s^{\mu\nu}$ and
$t^{\kappa\lambda\mu\nu}$ inherit the Ricci and Riemann
properties, respectively, and are traceless: $s^{\mu}_{{\phantom
\mu} \mu}=0$,
$t^{\kappa\lambda}_{\phantom{\kappa\lambda}\kappa\lambda}=0$,
$t^{\kappa}_{\phantom{\kappa}\mu\kappa\lambda}=0$ . The PPN
approximation of (\ref{LSaction}) has been studied in \cite{PPN}.
We restrict to the case $u=0$ and $t^{\kappa\lambda\mu\nu}=0$,
therefore only the coefficients $s^{\mu\nu}$ control the Lorentz
violation degrees of freedom. The variation of the action $S$ with
respect to the background metric yields the field equations
\cite{kosteleckyG} \vspace{-0.1in}
\begin{equation}\label{fieldequations}
  G^{\mu\nu}-(T^{Rs})^{\mu\nu}=8\pi G\, T_g^{\mu\nu}\,,
\end{equation}
where $G^{\mu\nu}=R^{\mu\nu}-(R/2)g^{\mu\nu}$ is the standard
Einstein tensor, and
\begin{widetext}
\begin{equation}\label{TRs}
 (T^{Rs})^{\mu\nu}=\frac{1}{2}\left(s^{\alpha\beta}R_{\alpha\beta}-\nabla_\alpha
                    \nabla_\beta\right) g^{\mu\nu}-s^{\mu\alpha}R_\alpha^{\phantom{\alpha}\nu}
                    -s^{\nu\alpha}R_\alpha^{\phantom{\alpha}\mu}
+\frac{1}{2}\left(\nabla_\alpha \nabla^\mu s^{\alpha
\nu}+\nabla_\alpha \nabla^\nu s^{\alpha \mu}\right)
-\frac{1}{2}\nabla^2 s^{\mu\nu}\,.
\end{equation}
\end{widetext}
Tracing (\ref{TRs}), one gets
\begin{equation}\label{traccia}
 R-\nabla_\alpha \nabla_\beta s^{\alpha \beta}=-8\pi G\, T_g\,,
\end{equation}
where $T_g=g_{\mu\nu}T_g^{\mu\nu}$. Moreover, from Bianchi's
identity $\nabla_\mu G^{\mu\nu}=0$, one obtains the relation
\begin{equation}\label{BianchiId}
  8\pi G\nabla_\mu T_{g\,\, \nu}^{\mu}=-\frac{1}{2}\, R^{\alpha\beta}\nabla_\nu s_{\alpha\beta}
        +R^{\alpha\beta}\nabla_\beta s_{\alpha\nu}+\frac{1}{2}\,s_{\alpha \nu}\nabla^\alpha R\,.
\end{equation}
The Friedman-Robertson-Walker (FRW) metric with zero spatial
curvature ($k=0$) is
\begin{equation}\label{FRWmetric}
 ds^2=dt^2-a^2(t)[dx^2+dy^2+dz^2]\,.
\end{equation}
We assume that the coefficients $s^{\mu\nu}$ preserve the FRW
symmetries (the Universe is isotropic and homogenous), so that
$s^{ij}=s^{00}\delta^{ij}/3$. After lengthly calculations, the 00
and $ij$ components of (\ref{fieldequations}), and Eq.
(\ref{BianchiId}) read
\begin{equation}
 3\frac{{\dot a}^2}{a^2}+\frac{\dot a}{a} {\dot s}^{\,00}-3\left(\frac{\ddot a}{a}-\frac{{\dot a}^2}{a^2}\right)
            s^{00} = 8\pi G \rho\,, \label{Eq00}
 \end{equation}
 \begin{equation}
 \frac{\ddot a}{a}+\frac{{\dot a}^2}{2a^2}+\left[\frac{{\ddot s}^{\,\,00}}{6}+\, \frac{\dot a}{a}{\dot s}^{\,00}
+\left(\frac{7}{6}\frac{\ddot a}{a}+\frac{5}{6}\frac{{\dot
a}^2}{a^2}\right)s^{00}\right] = -8\pi G p \label{Eqij}
 \end{equation}
   \[
  8\pi G\left[\frac{1}{a^3}\frac{\partial}{\partial t}(\rho a^3)+3p\,\frac{\dot a}{a}\right] =
   -\left(2\, \frac{{\ddot a}}{a}+\frac{{\dot a}^2}{a^2}\right){\dot s}^{\,00}
   \]
 \begin{equation} \label{ConsEn}
   -\left[3\, \frac{\dddot a}{a}+7\, \frac{\ddot a}{a}\frac{\dot a}{a}+2\left(\frac{\dot a}{a}\right)^3\right]s^{00}\,,
\end{equation}
where $\rho=T_{g\,\,0}^0$ and $p=T_{g\,\,i}^i$ (no sum over $i$).
Eq. (\ref{traccia}) reads
 \begin{equation}\label{tracciaFRW}
 R={\ddot s}^{\,\, 00}+7\, \frac{\dot a}{a}{\dot s}^{\, 00}+4\left(\frac{\ddot a}{a}+2\frac{{\dot a}^2}{a^2}\right)s^{00}
-8\pi G(\rho-3p)\,.
 \end{equation}
Notice that the energy conservation (\ref{ConsEn}) is fulfilled
for $\rho$ and $p$ given by (\ref{Eq00}) and (\ref{Eqij}).

In the conventional General Relativity, the pressure $p$ and
density $\rho$ are related, during the radiation era, by the
relation $p=\rho/3$ ($w=1/3$). From the Einstein field equations
it follows that the scale factor depends on cosmic time as
$a(t)\sim t^{1/2}$, the matter density as
$\rho_R(t)=\frac{3}{32\pi G}t^{-2}\sim a^{-4}$, and finally the
temperature as $T_R^2(t)=\frac{3}{4\pi}\sqrt{\frac{5}{G
g_*\pi}}t^{-1}\sim a^{-2}$. In the case of SME with gravity, the
constant $w=1/3$ is corrected by coefficients breaking the Lorentz
symmetry, and therefore may vary with the time $t$.

As discussed by Colladay and McDonald \cite{statistics}, the
formalism of statistical mechanics (as well as the laws of
thermodynamics) in presence of Lorentz breakdown is the same as
for conventional statistical mechanics. We therefore use the
standard definition for the density $\rho$ and the pressure $p$.
The general expressions for them are
 \begin{equation}\label{rho}
 \rho=g\int \frac{d^3P}{(2\pi)^3}\frac{u_\mu u_\nu P^\mu P^\nu}{P^0}f\,,
 \end{equation}
 \begin{equation}\label{p}
 p=\frac{g}{3}\int \frac{d^3P}{(2\pi)^3}\frac{(u_\mu u_\nu - g_{\mu\nu})P^\mu P^\nu}{P^0}f\,,
 \end{equation}
with $u^\mu=(1,0,0,0)$ the four-velocity of the fluid,
$f=(e^{E/T}\pm 1)$ is the Fermi-Dirac/Bose-Einstein distribution.
The dispersion relation reads $g_{\mu\nu}P^\mu P^\nu
=m^2+2c_{\mu\nu} P^\mu P^\nu+\ldots$, where the ellipsis
represents other contributions of SME. $c_{\mu\nu}$ in the SME
with gravity may depend on the position. In the radiation era, the
Universe was filled by photons and ultra-relativistic fermion
particles (baryons, electrons, neutrinos). According to SME
\cite{kostelecky}, let us assume the photon sector as
conventional, so that coefficients for the Lorentz violation enter
in the dispersion relations of ultra-relativistic particles.
Without loss of generality, let us consider Lorentz violating
corrections only for the baryon sector, i.e. $c_{00}^{(b)}\neq 0$
(otherwise, $c_{\mu\nu}^{(b)}$ is replaced by $\sum_a q_a
c_{\mu\nu}^{(a)}$, where $a=e, \nu, b, \ldots$, and $q_a$ are
constants which accounts for the statistics of particles).
Requiring the FRW symmetry, the non vanishing (traceless)
coefficients $c_{\mu\nu} (\ll 1)$ for the Lorentz violation are
the diagonal components, with $c^{ij}=c^{00}\delta^{ij}/3$ (see
\cite{lambiaseBBN}). As a consequence, one gets $c_{\mu\nu}P^\mu
P^\mu\simeq 4c_{00}E^2/3$, $E\simeq (1-4c_{00}/3)P$, and the
massive term turns out to be modified as $m^2(1+c^{00})$. The
latter may be neglected for ultra-relativistic particles ($P^2
\ggg m^2$). Eq. (\ref{p}) then reads
\begin{equation}\label{p1}
  p = \frac{\rho}{3}\left(1-\frac{7}{3}\frac{g_b}{g_*}c_{00}^{(b)}\right)\,.
\end{equation}
To account for corrections induced by $c^{00}$ and $s^{00}$
coefficients, also the scale factor turns out to be modified, and
therefore we set $a\to a(1+\delta)$, with $\delta \ll 1$. A
solution of Eqs. (\ref{Eq00})-(\ref{ConsEn}) can be derived with
the ansatz
 \[
a\simeq t^\alpha\,\quad  s^{00}=St^\gamma\,, \quad \delta =
Dt^\gamma\,, \quad
 c_{00}^{(b)}=Ct^\gamma\,,
 \]
To leading order in $s^{00}$, $c_{00}^{(b)}$, and $\delta$, and
using (\ref{p1}), Eqs. (\ref{Eq00})-(\ref{ConsEn}) admit the
following solution:
\begin{eqnarray}
 \gamma&=&-\frac{3}{2}\,, \quad \alpha=\frac{1}{2}\,, \label{gamma}\\
 \rho &=&
 \rho_R+\frac{8\gamma S+\frac{21}{2}\frac{g_b}{g_*}C}{32\pi
 G(3\gamma+1)}\frac{1}{t^{2-\gamma}}\,, \label{rhofinal}\\
 p &=& \frac{\rho_R}{3}+\frac{8\gamma S+\frac{7(1-6\gamma)}{2}\frac{g_b}{g_*}C}{96\pi G(3\gamma+1)}
 \frac{1}{t^{2-\gamma}}\,, \label{pfinal}
 \\
 D &=&
 \frac{-1}{2\gamma(\gamma+1)}\left[(\gamma+2)\left(\gamma+\frac{1}{2}\right)S+\frac{7}{4}\frac{g_b}{g_*}C\right]
\end{eqnarray}
The constants $S$ and $C$ are free and their combination is fixed
by the observed baryon number asymmetry. Moreover, Eqs.
(\ref{rhofinal}) and (\ref{pfinal}) indicate that corrections
induced by Lorentz violating terms fall down faster than
$\rho_R\sim 1/t^2$.

The interaction (\ref{riccicoupling}) generates a net baryon
asymmetry provided ${\dot R}\neq 0$. $s^{00}$ and
$c_{00}^{(b)}$-corrections prevent indeed the Ricci curvature to
vanish, as well as its first time derivative. From
(\ref{tracciaFRW}) it follows in fact
\begin{equation}
 {\dot
 R}=-\frac{(1-\gamma/2)\Pi}{t^{3-\gamma}}\,,
\end{equation}
where
 \[
\Pi\equiv \left|S+\frac{7g_b}{2g_*}C\right|\,.
 \]
The temperature is derived by using Eqs. (\ref{rho}) and
(\ref{rhofinal}), and is given by
$T^4(t)=T^4_R(t)+\delta_T/t^{2-\gamma}$, where $\delta_T$ includes
the $C$ and $S$-corrections. In computing $n_B/s$ (Eq.
(\ref{nB/s})), $\delta_T$ may be neglected since gives rise to
corrections of the second order. The net baryon asymmetry at the
decoupling temperature $T_D$ is hence
 \begin{equation}\label{basymfin}
 \frac{n_B}{s}\simeq 1.1\,\, 10^5 \Pi GeV^{-\gamma}\left(\frac{T_D}{m_{Pl}}\right)^{5-\gamma}
 \left(\frac{T_D}{GeV}\right)^{-\gamma}\left(\frac{m_{Pl}}{M_*}\right)^2
\end{equation}
where $m_{Pl}\sim 10^{19}GeV$ is the Planck mass. As pointed out
in \cite{steinhardt}, a possible choice of the cutoff scale is
$M_*=m_{Pl}$ if $T_D=M_I$, where $M_I\sim 3.3 \,\,10^{16}GeV$ is
the upper bound on the tensor mode fluctuation constraints in
inflationary scale. This choice is particular interesting because
implies that tensor mode fluctuations should be observed in the
next generation of experiments. Since the constraint of the
observed baryon asymmetry is $n_B/s\lesssim 9\,\, 10^{-11}$, one
gets
 \begin{equation}\label{PiFinal}
 \Pi\lesssim 5.8 \,\, 10^{-24.5} GeV^{-3/2}\sim 3.1\,\, 10^{-60.5} sec^{3/2}\,,
 \end{equation}
An upper bound for $s^{00}$ is therefore
 \begin{equation}\label{s00Final}
 s^{00} \lesssim 7.6\, 10^{4}\left(\frac{t_{Pl}}{t}\right)^{3/2}
 \quad for \quad t \gtrsim 10^{-38}sec\,.
 \end{equation}
where $t_{Pl}\sim 5.4\,\,10^{-44}sec$ is the Planck time.
Similarly for $c_{00}^{(b)}$:
 \begin{equation}\label{c00Final}
 c^{(b)}_{00} \lesssim 1.1\, 10^{6}\left(\frac{t_{Pl}}{t}\right)^{3/2}
 \quad for \quad t \gtrsim 10^{-38}sec\,.
 \end{equation}
As we can see, coefficients leading to deviations from General
Relativity, as well as to Lorentz's and CPT violation, decrease
during the expanding phase of the Universe, and therefore
corrections become more and more negligible. Of course such
results refer to radiation dominated era. Different epochs, such
as, for example, matter dominated era, lead to different field
equations, hence to a different time-dependence of the Lorentz and
CPT violating coefficients, with potentially interesting
consequences on cosmological scenarios offered by the SME with
gravity.


In conclusion, a cosmological consequence of the gravitational
sector in the SME has been investigated in this paper. Such a
study is related to the (CPT violating) interaction between the
Ricci scalar curvature and the baryon current (Eq.
(\ref{riccicoupling})). During the phase of the expanding Universe
dominated by radiation, SME with the inclusion of gravity provides
a framework in which the baryon asymmetry may be gravitationally
induced. The current estimations on the observed baryon asymmetry
yield an order of magnitude of coefficients breaking the Lorentz
and CPT symmetry.


\vspace{0.1in}

\acknowledgments It is a pleasure to thank V. Alan Kosteleck\'y
for comments and suggestions.


\begin{thebibliography}{99}
\bibitem{note} Actually, the current $J^\mu$ can
be replaced by any current yielding a net $B-L$ charge in thermal
equilibrium ($B$ and $L$ are the baryon and lepton numbers,
respectively), so that the asymmetry is not wiped out by the
electroweak anomaly \cite{kuzmin}.
\bibitem{note1} It is assumed that the
B-violating processes are generated by an operator of mass
dimension $D=4+n$ \cite{steinhardt}, which decouple to $T_D$. The
interaction rate is $\Gamma_B = T^{2n+1}/M_B^{2n}$, where $M_B$ is
the mass scale associated with the operator that generates
B-violating proccesses.
\end{thebibliography}
\end{document}